\documentclass[9pt,twocolumn,twoside]{opticajnl}
\journal{opticajournal} 

\setboolean{shortarticle}{true}

\usepackage{graphicx}
\usepackage{caption}
\usepackage{siunitx}

\newcommand{\ea}{\textit{et al.}}
\newcommand{\omegahf}{\Omega_{\rm HF}}
\newcommand{\omegalf}{\Omega_{\rm LF}}

\DeclareSIUnit\dbm{dBm}

\usepackage{lineno}

\title{Digital stabilization of an I/Q modulator in the carrier suppressed single side-band (CS-SSB) mode for atom interferometry}

\author[1]{Arif Ullah}
\author[1]{Samuel Legge}
\author[1]{John D. Close}
\author[1]{Simon A. Haine}
\author[1,*]{Ryan J. Thomas}

\affil[1]{Department of Quantum Science and Technology, Research School of Physics, The Australian National University, Canberra 2601, Australia}

\affil[*]{ryan.thomas@anu.edu.au}

\begin{abstract}
We present an all-digital method for stabilising the phase biases in an electro-optic I/Q modulator for carrier-suppressed single-sideband modulation.  Using programmable logic on the Red Pitaya STEMlab 125-14 platform, we digitally generate and demodulate an auxiliary radio-frequency tone whose beat with the optical carrier probes the I/Q modulator's phase imbalances.  We implement a multiple-input, multiple-output integral feedback controller which accounts for unavoidable cross-couplings in the phase biases to lock the error signals at exactly zero where optical power fluctuations have no impact on phase stability.  We demonstrate $>23\,\rm dB$ suppression of the optical carrier relative to the desired sideband at $+3.4\,\rm GHz$ over a period of $15$ hours and over temperature variations of $20^\circ\rm C$.
\end{abstract}

\setboolean{displaycopyright}{false} 

\begin{document}

\maketitle

\section{Introduction}

Many quantum technologies, such as sensors \cite{kasevich_measurement_1992,peters_measurement_1999,freier_mobile_2016,bidel_absolute_2018,bidel_airborne_2023}, memories \cite{phillips_storage_2001,appel_quantum_2008,lvovsky_optical_2009,hosseini_unconditional_2011}, and computers \cite{bluvstein_quantum_2022,levine_dispersive_2022,bluvstein_logical_2024}, rely on the coherent manipulation of quantum states of neutral atoms using two or more lasers at different frequencies which have a stable phase relationship while simultaneously being tunable \cite{templier_carrier-suppressed_2021}, spectrally pure \cite{carraz_phase_2012}, and high power \cite{lee_compact_2022,sane_11_2012,templier_carrier-suppressed_2021}.  In the case of systems using alkali metal atoms, such as rubidium and cesium, these frequencies are often separated by several GHz corresponding to the ground-state hyperfine splitting \cite{kasevich_atomic_1991,hosseini_high_2011,bluvstein_quantum_2022}, which complicates coherent frequency synthesis.  Quantum sensors and memories may also be deployed in the field \cite{wang_field-deployable_2022,menoret_gravity_2018,wu_gravity_2019,freier_mobile_2016} which further requires the laser system to be robust to environmental changes and, particularly for quantum sensors, to have low size, weight, and power (SWaP) \cite{lee_compact_2022,bidel_airborne_2023,kodigala_high-performance_2024}.

One method for generating phase-coherent frequencies at GHz separations is to modulate the phase of the laser using an electro-optic modulator (EOM) \cite{theron_narrow_2015,menoret_gravity_2018,li_novel_2020}.  Sinusoidal modulation of the phase yields sidebands symmetrically located on either side of the carrier frequency at multiples of the modulation frequency.  By interferometrically combining the output laser fields from several EOMs, usually in a single fibre-coupled waveguide structure, unwanted spectral components can be suppressed.  In particular, the so-called dual-parallel Mach-Zehnder interferometer (MZI) structure creates an I/Q modulator that allows independent control of the amplitude and phase of the light \cite{izutsu_integrated_1981}.  Careful tuning of the phase biases of the internal MZIs allows the modulator to operate in the carrier-suppressed single sideband (CS-SSB) mode where nearly all of the optical power is in one of the sidebands.  As these phase biases are prone to drift, especially due to temperature but also due to photorefractive damage \cite{hall_photorefractive_1985,kong_recent_2012}, they need to be actively stabilised at the correct operating point.  The most common scheme involves modulating the phase biases at the \si{\kilo\hertz} level \cite{bui_instrumentation_2011,templier_carrier-suppressed_2021} which is transferred to the optical signal and then detected and demodulated to form an error signal.  The disadvantage of this method is that the output laser power now varies at the modulation frequency, which can lead to pulse area errors in the manipulation of quantum states.

An alternative stabilisation method was demonstrated by Wald \ea~\cite{wald_analog_2023} in which a weak auxiliary radio-frequency (RF) tone is added to the main microwave tone.  Demodulation of the measured optical power at the RF tone and its second harmonic yields signals that measure the amplitude of the carrier and unwanted sideband.  When operating in CS-SSB mode only one sideband from each frequency is present, and the beatnote between these frequencies is on the order of GHz, eliminating the effect of amplitude modulation on the manipulation of quantum states.  Stabilisation of the phase biases was achieved using analog electronics and three independent feedback loops; however, cross-coupling between the interferometer phase biases meant that the error signals had to be stabilised away from zero.  This leads to the possibility that the spectral purity of the output mode can be compromised by changes in the input laser power or in the environment.  

Building on the work presented in \cite{wald_analog_2023}, we present an all-digital system for stabilising the phase biases in an I/Q modulator for CS-SSB operation.  We use the fast analog input and output capabilities of the Red Pitaya STEMlab 125-14 platform, combined with custom high-speed programmable logic, to implement a digital multiple-input, multiple-output (MIMO) feedback loop that compensates for cross-coupling of the phase biases and allows locking of the measured signals at exactly zero.  Direct access to the system's transfer function allows us to automatically and efficiently tune the feedback loop.  We demonstrate stable operation of our system over long periods of time (\SI{15}{\hour}) and large temperature variations (\SI{20}{\celsius}).  The entire laser system is fibre-coupled, with no free space components, providing a robust and compact coherent frequency source required for field-deployable quantum technologies such as atom interferometer inertial sensors.

\section{Carrier-suppressed single sideband generation}\label{Sec:IQmod}

\begin{figure}[t]
   \centering
    \includegraphics[width=\columnwidth]{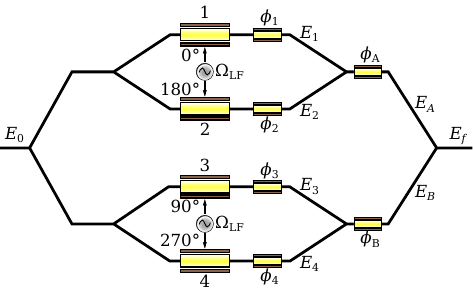}
    \caption{Schematic of the dual-parallel MZI I/Q modulator, with optical paths shown as black lines. High frequency phase modulators are labelled 1-4 and are driven at $\omegalf$ with noted phases. Various electric fields ($E_x$) and bias phase control points ($\phi_x$) are described in Section \ref{Sec:IQmod}.}
     \label{fig:IQ}
\end{figure}

We consider an I/Q modulator in the dual parallel Mach-Zehnder interferometer (MZI) configuration as shown in Fig.~\ref{fig:IQ} with a single modulation frequency $\omegalf$. A field of amplitude $E_0$ with carrier frequency $\omega_c$ is applied to the I/Q modulator and is split into four initial paths, each of which goes through a phase modulator. These phase modulators are arranged in a pair of MZIs, with one pair (1-2) being driven by signals $\pm\sin\omegalf t$ and the other (3-4) being driven by $\pm\cos\omegalf t$.  In addition to the modulation, we also apply phases $\phi_1$, $\phi_2$, $\phi_3$, and $\phi_4$ to each path.  The output fields $E_A$ and $E_B$ of the sub-MZIs are subject to additional phase shifts $\phi_A$ and $\phi_B$ and are then passed through the outer MZI to get the final output field $E_f$. 

We consider a phase modulation of the form $\beta \sin \omegalf t$ with an amplitude $\beta \ll 1$ such that 
\begin{equation}
    \exp[i\beta \sin (\omegalf t)]\approx 1 + \frac{\beta}{2}  e^{i\omegalf t}-\frac{\beta}{2} e^{-i\omegalf t}.
\end{equation}
Defining average phases $\bar{\phi}_A = \frac{\phi_1+\phi_2}{2}$, $\bar{\phi}_B = \frac{\phi_3+\phi_4}{2}$, and $\bar{\phi} = \frac{\bar{\phi}_A + \bar{\phi}_B}{2}$, as well as differential phases $\delta_A = \frac{\phi_1-\phi_2}{2}+\frac{\pi}{2}$, $\delta_B = \frac{\phi_3 -\phi_4}{2}+\frac{\pi}{2}$, and $\delta_P = \frac{\phi_A-\phi_B}{2} + \frac{\bar{\phi}_A - \bar{\phi}_B}{2}$, the field at the output of the I/Q modulator is
\begin{align}
    E_f &= -\frac{E_0 e^{i\omega_c + i\bar{\phi}}}{2} \Bigg[\left(e^{i\delta_P} \sin \delta_A + e^{-i\delta_P} \sin \delta_B \right) \nonumber \\
        & + \frac{\beta}{2} e^{i\omegalf t} \left(-i e^{i\delta_P} \cos \delta_A + e^{-i\delta_P} \cos \delta_B \right) \nonumber\\
        & + \frac{\beta}{2} e^{-i\omegalf t} \left(i e^{i\delta_P} \cos \delta_A + e^{-i\delta_P} \cos \delta_B \right)\Bigg].
    \label{eq:final-field}
\end{align}
The carrier is suppressed when $\sin\delta_A = \sin\delta_B = 0$, and by setting $\delta_P = \pm\pi/4$ the positive or negative sideband can be selected to give CS-SSB operation.  The relative amplitudes of the various sidebands can be determined by measuring the total power $P \propto |E_f|^2$ using a photodetector with a bandwidth $\gg 2\omegalf$ and is
\begin{align}
    |E_f|^2 &= \frac{|E_0|^2}{4} \Bigg[\sin^2 \delta_A + \sin^2 \delta_B + 2 \sin \delta_A \sin \delta_B \cos 2\delta_P \nonumber\\
    & + \frac{\beta^2}{2} (\cos^2 \delta_A + \cos^2 \delta_B) \nonumber\\
    & + 2 \beta \cos \delta_A (\sin \delta_A + \cos 2\delta_P \sin \delta_B) \sin \omegalf t \nonumber\\
    & + 2 \beta \cos \delta_B (\sin \delta_B + \cos 2\delta_P \sin \delta_A) \cos \omegalf t \nonumber\\
    & + \frac{\beta^2}{2} \left(2 \cos \delta_A \cos \delta_B \cos 2\delta_P \sin 2 \omegalf t\right. \nonumber\\
    & + \left.(\cos^2 \delta_B - \cos^2 \delta_A) \cos 2 \omegalf t\right)\Bigg]
    \label{eq:power-full}
\end{align}
where the first two lines are the DC power in the different frequency components, the next two lines are the beatnote between the carrier and $\pm1$ sidebands, and the final two lines are the beatnote between the $\pm1$ sidebands.  Near the CS-SSB operating point, where $\delta_A,\delta_B\approx 0$ and $\delta_P \approx \pm\pi/4$, we have
\begin{align}
    |E_f|^2 &\approx |E_0|^2\left[\frac{\beta^2}{4} + \frac{\beta}{2}\delta_A\sin\omegalf t + \frac{\beta}{2}\delta_B\cos\omegalf t\right.\nonumber\\
    &  \left.+ \frac{\beta^2}{2}\left(\delta_P \mp \frac{\pi}{4}\right)\sin 2\omegalf t\right]
    \label{eq:power-approx}
\end{align}
which implies that the amplitudes of the photodetector signals at $\omegalf$ and $2\omegalf$ are linear in the phase biases and thus can be used as error signals for locking the phase biases at the correct operating points \cite{wald_analog_2023}.  Note that, according to \eqref{eq:power-full}, if $\cos\delta_A = \cos\delta_B = 0$, which corresponds to carrier-only operation, our error signals will also be near zero and linear in phase deviations about that point, so for this locking scheme to work it needs to start close to the desired CS-SSB set-point by using auxiliary information such as known voltage biases for the I/Q modulator or an optical spectrum analyzer.

\begin{figure*}[t]
	\centering
    \includegraphics[width=\textwidth]{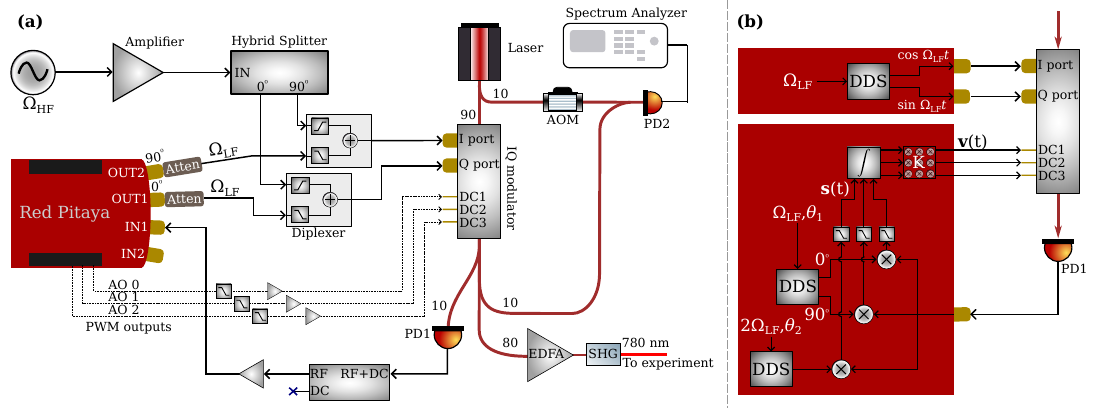}
    \caption{\textbf{(a)} Experimental setup with optical fibre paths shown as red lines and electrical paths shown in black lines.  The initial laser beam passes through a fibre beam-splitter with $90\%$ used as the input to the I/Q modulator, while the remainder passes through an AOM and is used for measurement of the beat note signal by beating it with the I/Q modulator output beam.  High frequency $\omegahf$ modulation from a signal generator passes through a $90^\circ$ hybrid splitter and is combined with the low frequency $\omegalf$ signals from a Red Pitaya STEMlab 125-14 (RP) using diplexers to drive the I/Q modulator.  After I/Q modulation, $10\%$ of the power is detected on photodetector PD1, amplified, and measured using one of the fast analog inputs on the RP, $10\%$ is used for out-of-loop monitoring on PD2, and the remaining $80\%$ is amplified and frequency-doubled using a waveguide second-harmonic generation (SHG) crystal. The I/Q modulator's phase biases (DC 1, 2, and 3) are controlled using suitably filtered and amplified pulse-width modulation (PWM) outputs AO 0-2.  \textbf{(b)} Simplified block diagram illustrating the core feedback loop and digital logic within the Red Pitaya FPGA.}
    \label{fig:setup}
\end{figure*}

\section{Experimental setup}

Our optical and electronic setup is shown in Fig.~\ref{fig:setup}a.  A narrow-linewidth fiber laser (NKT Photonics Koheras Basik) is used as a seed laser that generates \SI{30}{\milli\watt} of laser light at \SI{1560}{\nano\meter}.  The laser beam is then passed through a fiber beam-splitter where $90\%$ is used as an input to the I/Q modulator (iXblue MXIQ-LN-30), while the remainder is used as a source for a fibre-coupled acousto-optic modulator (AOM, G\&H 80MHz Fiber-Q). A microwave synthesizer generates a high frequency signal at $\omegahf = \SI{3.4}{\giga\hertz}$ which is first amplified and then passed through a $90^\circ$ hybrid power splitter (Mini Circuits ZX 100-2-34-S+) to get two signals with a phase difference of $90^\circ$ at a power of $\mathord{\sim}\SI{20}{\dbm}$.  We combine the microwave signals with two low-frequency ($\omegalf = \SI{4}{\mega\hertz}$) signals that are $90^\circ$ out of phase generated by the fast analog outputs of a Red Pitaya (RP) STEMlab 125-14 board (power of $\sim\!\SI{-17}{\dbm}$) using a pair of diplexers (Marki DPXN-OR5) which then also drive the two RF ports of the I/Q modulator.  The exact frequency for $\omegalf$ is unimportant so long as $\omegalf$ is much faster than the timescales associated with manipulating quantum states: we chose \SI{4}{\mega\hertz} to avoid a noise band for frequencies $<\SI{3}{\mega\hertz}$, and so that both $\omegalf$ and $2\omegalf$ are well-sampled by the RP's ADCs.  We found it was important to avoid multiples of \SI{5}{\mega\hertz}, as we observed small glitches in the demodulated signals which we traced to an unwanted interaction with the FPGA clock frequency at \SI{125}{\mega\hertz}.  Three pulse-width modulation (PWM) outputs from the RP with $10$ bit resolution clocked at \SI{250}{\mega\hertz} are amplified to the range of $[0,14]$ \si{\volt} and filtered with a corner frequency of $\mathord{\sim}\SI{10}{\hertz}$ to drive the three phase biases on the I/Q modulator.  

After the I/Q modulator, we split the laser using two fiber beam-splitters such that $10\%$ of the light impinges on the feedback photodetector (PD1, Thorlabs DET01CFC), $10\%$ is combined with the light that passes through the AOM and is measured with a fast photodetector (PD2, Thorlabs DET08C), and $80\%$ passes through an erbium-doped fiber amplifier (Civil Laser EDFA-C-BA-23-PM-M) which is then frequency doubled using a waveguide second harmonic generator (SHG, Covesion WGCO-H-1560-40) crystal to produce light at \SI{780}{\nano\meter} for our rubidium atom interferometry experiments \cite{templier_carrier-suppressed_2021,lee_compact_2022}. All fibre components use polarization maintaining fibre. 

The signal from PD1 passes through an RF bias-tee to remove the DC component and is then amplified by $\sim\!\SI{40}{\deci\bel}$ before being recorded using one of the fast analog inputs on the RP, and this serves as the source for feedback error signals.  With the AOM shifting the frequency of the carrier by \SI{80}{\mega\hertz}, the spectrum measured at PD2 can unambiguously measure the power in each of the frequencies produced by the I/Q modulator and allows for out-of-loop measurements of the system's performance.

\section{Digital stabilisation of phase biases}

\begin{figure}[tb]
    \centering
    \includegraphics[width=\columnwidth]{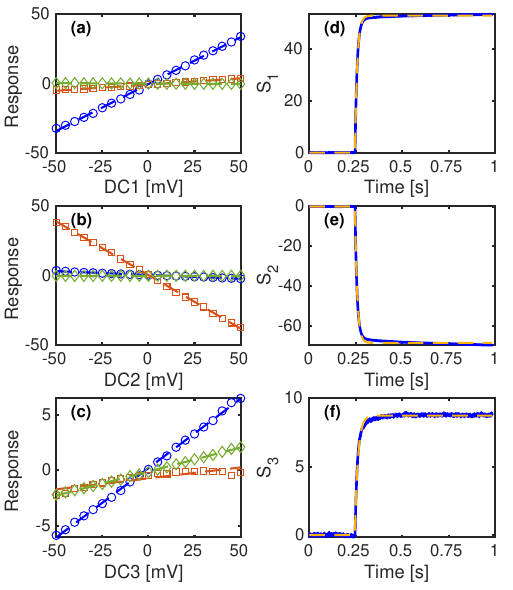}
    \caption{Measurement of the open-loop transfer function.  \textbf{(a-c)} Measured signals $s_1$ (blue circles), $s_2$ (red squares), and $s_3$ (green diamonds) for small changes in the DC voltages about the CS-SSB operating point.  Dashed lines are linear fits to the data. \textbf{(d-f)} Transient response of signals $s_1$, $s_2$, and $s_3$ to sudden jumps in voltages DC1, DC2, and DC3, respectively.  Dashed orange lines are exponential fits to the data (blue).}
    \label{fig:Optim}
\end{figure}

We use custom firmware \cite{github} running on the RP's on-board field-programmable gate array (FPGA) for digital stabilization of the phase biases of the I/Q modulator.  Digital feedback loops offer several advantages over analog feedback loops: set-points have no noise or drift, the system is both easily modified and highly portable to other experiments, and complex control laws are readily implemented.  The major disadvantages of digital loops are that they can suffer from digitisation noise on inputs and outputs, and they tend to have lower loop bandwidths due to the latency of digital operations.  However, for stabilising the phase biases neither of these potential drawbacks cause problems.  We amplify the input signal and use a sufficiently large PWM bit depth so that digitisation noise is not a limiting factor.  Changes in the phase biases are typically thermally-driven, with timescales on the order of \SI{100}{\milli\second}, which are long enough that digital latency (on the order of \SI{500}{\nano\second}) is a negligible contribution to the loop delay.  

Our implementation, which is illustrated in Fig.~\ref{fig:setup}b, uses three direct digital synthesis (DDS) modules, one of which drives the two analog outputs at $\sin\omegalf t$ and $\cos\omegalf t$.  The other two DDSs generate signals $\sin(\omegalf t + \theta_1)$, $\cos(\omegalf t + \theta_1)$, and $\sin (2\omegalf t + \theta_2)$ with demodulation phases $\theta_1$ and $\theta_2$ which are multiplied with the input signal and then filtered using third-order cascaded integrator-comb (CIC) filters to produce a $3\times 1$ vector of measurements $\mathbf{s}$.  The CIC filters also reduce the sampling rate so that the measurements can be more easily processed and stored for later analysis: we use a rate reduction of $2^{13}$ for a sample rate of $\mathord{\sim}\SI{15}{\kilo\hertz}$ which is much faster than the timescales at which the phase biases change.  From \eqref{eq:power-full} and \eqref{eq:power-approx}, the measurements are approximately
\begin{equation}
    \mathbf{s} \approx g|E_0|^2\begin{bmatrix}
        \beta(\delta_A\cos\theta_1 + \delta_B\sin\theta_1)\\
        \beta(-\delta_A\sin\theta_1 + \delta_B\cos\theta_1)\\
        \beta^2\left(\delta_P \mp \frac{\pi}{4}\right)\cos\theta_2 - \frac{\beta^2}{2}(\delta_A^2 - \delta_B^2)\sin\theta_2
    \end{bmatrix}
    \label{eq:measurements}
\end{equation}
with overall measurement scaling $g$.  Ideally, the phase biases can be adjusted independently, so by correctly choosing the demodulation phases (in this instance $\theta_1 = \theta_2 = 0$) we maximize the measurement amplitudes and minimize cross-coupling between measurements.  In practice, the DC voltages $\mathbf{v}$ used to change the phase biases couple significantly to the other biases, and it is impossible to entirely eliminate cross-coupling.  

To appropriately account for cross-coupling between $\mathbf{v}$ and $\mathbf{s}$, we use a discretized multiple-input multiple-output (MIMO) integral control law, so that $\mathbf{v}(t) = \mathrm{K}\int_{-\infty}^t[\mathbf{r} - \mathbf{s}(t')]dt'$ for static set-points $\mathbf{r}$ and $3\times 3$ feedback matrix $\mathrm{K}$.  We implement the control law in programmable logic using a time-shared multiplier to reduce system resources ($1$ multiplier instead of $9$), as the additional latency ($\sim\!\SI{250}{\nano\second}$) is negligible compared to other timescales.  We model the response of the measurements to changes in $\mathbf{v}$ near the CS-SSB operating point as a first-order low-pass filter with steady-state transfer matrix $\mathrm{G}$ and diagonal decay rate matrix $\Gamma$
\begin{equation}
    \dot{\mathbf{s}} + \Gamma\mathbf{s} = \Gamma\mathrm{G}\mathbf{v},
\end{equation}
where the decay rates $[\Gamma]_{ii}\approx 2\pi\times\SI{10}{\hertz}$ are dominated by the PWM low-pass filter rather than the inherent response of the I/Q modulator's phase bias ports or by the measurement filter response.  The closed-loop behaviour is then a second-order dynamical system
\begin{equation}
    \ddot{\mathbf{s}} + \Gamma\dot{\mathbf{s}} + \Gamma\mathrm{G}\mathrm{K}\mathbf{s} = \Gamma\mathrm{G}\mathrm{K}\mathbf{r}.
    \label{eq:closed-loop-response}
\end{equation}
By measuring $\Gamma$ and $\mathrm{G}$, we can choose a feedback matrix $\mathrm{K}$ such that $\Gamma\mathrm{G}\mathrm{K}$ is diagonal, which minimizes cross-coupling of measurement noise and environmental perturbations, and has eigenvalues close to those required for critical damping.  With critical damping, the loop bandwidth of the controller is $\Gamma$.  

We automatically tune our control system in several steps.  First, we determine the demodulation phase $\theta_2$ by fixing DC1 and DC2 and varying DC3 over its full range for demodulation phases $\theta_2 \in [0,\pi]$: the optimum value of $\theta_2$ is where the amplitude of $s_3$ as a function of DC3 is maximal as seen in \eqref{eq:power-full}.  At this demodulation phase, we then vary DC3 and locate the values at which $s_3 = 0$ which correspond to $\delta_P \approx \pm \pi/4$.  Picking the desired sideband, we then vary DC2 with DC1 and DC3 fixed for several demodulation phases $\theta_1 \in [0,\pi]$ and measure where the response of $s_2$ is a maximum as a function of DC2: this is the optimum phase $\theta_1$.  We individually measure $s_1$ versus DC1 and $s_2$ versus DC2 and find where $s_1=s_2 = 0$; here, one needs additional information because $s_1=s_2$ = 0 both when the carrier is suppressed and when both sidebands are suppressed.  After picking the correct operating point, we measure $\mathrm{G}$ by making small changes to each of the DC voltages and measuring the linear response of $\mathbf{s}$ in the steady state as shown in Figs.~\ref{fig:Optim}a-c.  We measure $\Gamma$ by making a step change in $\mathbf{v}$ and measuring the relaxation of $\mathbf{s}$ to its steady state as shown in Figs.~\ref{fig:Optim}d-f.  We then choose a diagonal target feedback matrix $\mathrm{K}_t$ such that $\mathrm{K} = \mathrm{G}^{-1}\Gamma^{-1}\mathrm{K}_t$; the elements of $\mathrm{K}_t$ are the squares of the natural frequencies of the decoupled harmonic oscillators in \eqref{eq:closed-loop-response}.  One then aims for $[\mathrm{K}_t]_{ii} = [\Gamma]^2_{ii}/4$ to achieve critical damping.

\section{Characterization}

\begin{figure}[tb]
    \centering
    \includegraphics[width=\columnwidth]{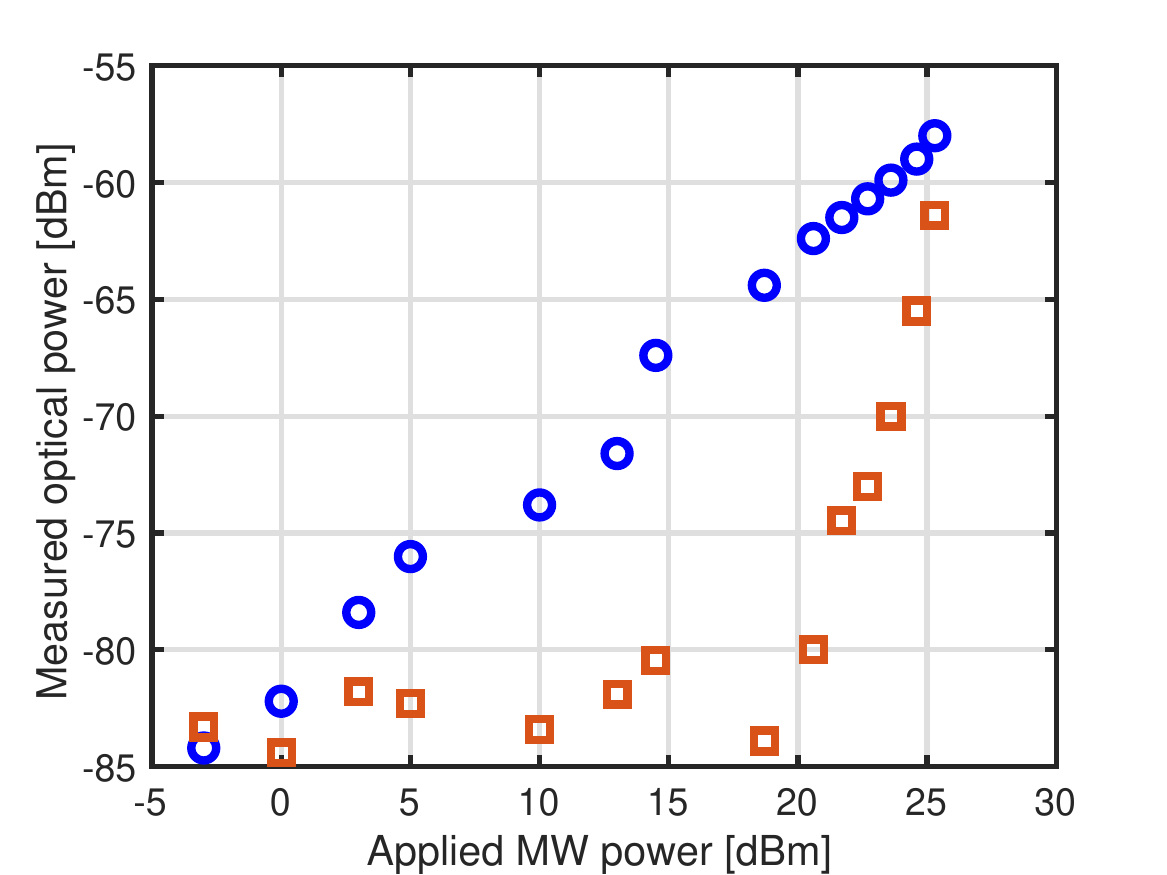}
    \caption{Carrier suppression as a function of applied microwave power at \SI{3.4}{\giga\hertz} when feedback is engaged.  Blue circles show the measured power in the positive sideband, while red squares show the measured power in the carrier.}
    \label{fig:carrier-suppression}
\end{figure}

\begin{figure}[tb]
    \centering
    \includegraphics[width=\columnwidth]{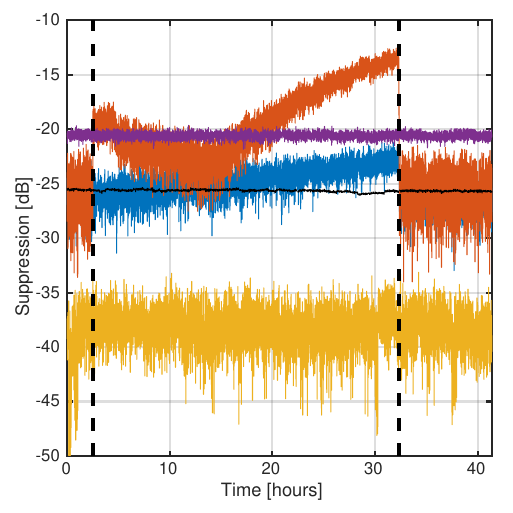}
    \caption{Stability of CS-SSB over time.  Suppression of carrier (red), $-\omegahf$ (blue), $+\omegalf$ (purple), and $-\omegalf$ (yellow) relative to the desired $+\omegahf$ sideband, measured using PD2 in Fig.~\ref{fig:setup}a.  Black horizontal line indicates the lower measurement bound on the carrier due to the AOM, whereas the black vertical dashed lines separate the locked and unlocked data.}
    \label{fig:time-series}
\end{figure}

\begin{figure}[tb]
    \centering
    \includegraphics[width=\columnwidth]{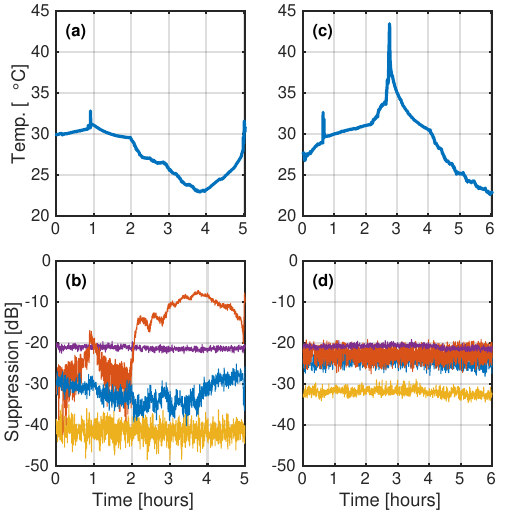}
    \caption{Suppression of unwanted spectral components relative to desired $+\omegahf$ sideband, measured using PD2 in Fig.~\ref{fig:setup}a, when the I/Q modulator is subjected to large temperature variations.  Suppression of carrier (red), $-\omegahf$ (blue), $+\omegalf$ (purple), and $-\omegalf$ (yellow), recorded over a period of 5 and 6 hours as shown in bottom panels for the unlocked \textbf{(b)} and locked modulator \textbf{(d)}, by changing the modulator temperature as shown in top panels over the same period for the unlocked \textbf{(a)} and locked case \textbf{(c)}.}
    \label{fig:temperature-data}
\end{figure}

In order to investigate the stability of the I/Q bias lock and to characterise the quality of CS-SSB operation, we use the out-of-loop measurement scheme shown in Fig.~\ref{fig:setup}a which overlaps the output of the I/Q modulator with a frequency-shifted copy of the carrier and measures the resulting signal using a fast photodiode and spectrum analyzer.  We first measure the suppression of the carrier relative to the positive sideband for different applied microwave powers when the lock is engaged as shown in Fig.~\ref{fig:carrier-suppression}.  We find that the sideband power increases linearly with microwave power over the entire investigated range, implying that we are operating in a phase modulation regime where we can neglect higher modulation sidebands.  However, the carrier power is constant only for low microwave powers $P_{\rm MW} < \SI{20}{\dbm}$ and increases as $P_{\rm MW}^4$ otherwise.  Further investigation has shown that the combined presence of signals at $\omegalf$ and $\omegahf$ leads to the generation of additional $\omegalf$ amplitude on the optical signal which interferes with the $\omegalf$ signal used for bias stabilization, likely due to non-linearities in the phase modulators \cite{kodigala_high-performance_2024}.  This leads to the condition $\mathbf{s} = \mathbf{0}$ no longer being the CS-SSB operating point.  Choosing $\mathbf{r} \neq \mathbf{0}$ allows us to reduce the measured carrier power, but this raises the possibility that the lock stability now depends on the input optical power, insertion loss, and modulation depth which can change with environmental conditions or with time \cite{hall_photorefractive_1985,kong_recent_2012}.  Instead, we reduce the microwave power to \SI{20}{\dbm} so that we can continue operating with $\mathbf{r} = \mathbf{0}$ at the cost of a factor of two in sideband power, which in our case can be compensated with an optical pre-amplifier.

Figure \ref{fig:time-series} presents a time-series measurement illustrating the difference in stability between the locked and unlocked states.  When the biases are unlocked there is significant drift in the carrier suppression over the course of about $1$ day with a much smaller change in the suppression of the $-\omegahf$ sideband.  The timescale for variations in the carrier suppression is $\mathord{\sim}\SI{20}{\minute}$.  The power in the $\pm\omegalf$ sidebands remains approximately constant.  When the biases are locked, the carrier and $-\omegahf$ sideband are stabilised to powers that are on average \SI{-26}{\deci\bel} below the positive sideband.  This suppression is likely limited by noise on the measurements $\mathbf{s}$, corresponding to noise on the $\omegalf$ signal within the CIC filter bandwidth of $\mathord{\sim}\SI{15}{\kilo\hertz}$. This measurement noise is likely from the photodetector and amplifiers used to measure the optical power, as the expected noise contributed by the RP ADC is a factor of $4$ lower than what is measured.  The noise on the suppression measurements in Fig.~\ref{fig:time-series} is partially due to the noise floor of the spectrum analyzer, and partially due to real fluctuations in the suppression caused by the locking bandwidth of $\mathord{\sim}\SI{10}{\hertz}$; additional noise is present on the carrier measurement because of interference with the AOM driving signal used for this heterodyne measurement.

Figure \ref{fig:temperature-data} demonstrates time-series measurements of the suppressed carrier and unwanted sidebands with large temperature swings that we induce with a heat gun and by changing the temperature of the laboratory.  We observe that the locked system remains stable against large temperature variations over many hours, whereas the unlocked system displays large excursions from the CS-SSB operating mode.  Stability against temperature variations is especially critical for field-deployable quantum technologies such as quantum sensors or repeaters which may be subjected to large changes in their operating environment. 

\begin{figure}[t]
    \centering
    \includegraphics[width=\columnwidth]{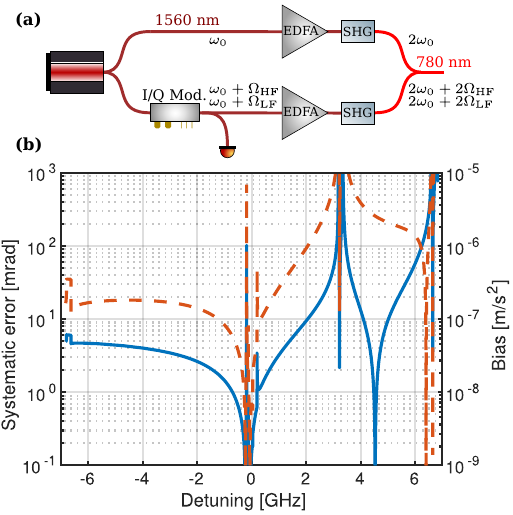}
    \caption{\textbf{(a)} Raman laser system with one arm subject to CS-SSB and one not, with primary frequencies before and after doubling labelled.  \textbf{(b)} Maximum systematic error for a Raman atom interferometer assuming the ratio of optical powers in the carrier and positive sideband frequencies are chosen to minimise the differential AC Stark shift.  Solid blue curve corresponds to the locked performance with suppressions of \SI{-25.9}{\deci\bel}, \SI{-26.3}{\deci\bel}, \SI{-20.6}{\deci\bel}, and \SI{-38.7}{\deci\bel} for the carrier, $-\omegahf$, $+\omegalf$, and $-\omegalf$ frequencies, respectively.  Dashed red curve corresponds to the unlocked performance with suppressions of \SI{-10}{\deci\bel} and \SI{-22}{\deci\bel} for the carrier and $-\omegahf$ sidebands, respectively, and unchanged suppressions for the $\pm\omegalf$ sidebands.  Acceleration bias (right axis) assumes an interferometer time of \SI{75}{\milli\second}.}
    \label{fig:phase-error}
\end{figure}

Stable, high suppression of unwanted frequency components is especially critical for atom interferometry, as the additional spectral components can cause large systematic phase shifts through longitudinal variation in the phase of the two-photon Raman Rabi frequency \cite{carraz_phase_2012,templier_carrier-suppressed_2021}.  In our experiment, the I/Q modulator path drives one of the lasers in the Raman transition, while the other laser is generated by amplifying and doubling the seed laser but with no frequency modifications to produce a \SI{780}{\nano\meter} laser at the same frequency as the carrier from the I/Q modulator \cite{kodigala_high-performance_2024}: see Fig.~\ref{fig:phase-error}a.  Adapting the model presented in Ref. \cite{templier_carrier-suppressed_2021}, and assuming that the ratio of optical powers in the carrier frequency and positive sideband frequency are chosen to minimise the differential AC Stark shift \cite{freier_atom_2017}, we calculate the maximum systematic error that would be seen with our laser system in Fig.~\ref{fig:phase-error}b.
We find that the unwanted sidebands contribute at most \SI{10}{\milli\radian} (\SI{e-7}{\meter/\second^2} for a $T = \SI{75}{\milli\second}$ interferometer) to the systematic error for Raman laser detunings from \SI{-6.8}{\giga\hertz} to \SI{+2}{\giga\hertz}, while the unlocked system can reach systematic errors of up to \SI{100}{\milli\radian} (\SI{e-6}{\meter/\second^2}) in that same region when ignoring detunings close to zero.

\section{Conclusions}

In this article, we presented a laser system based on digital stabilization of an I/Q modulator in the CS-SSB mode for state preparation and manipulation in quantum technologies and especially atom interferometers for inertial sensing.  The phase biases of the I/Q modulator were measured and stabilized by an all-digital feedback circuit utilizing the programmable logic and fast analog input/output capabilities of the Red Pitaya STEMlab 125-14 platform.  Careful application of microwave power and the use of a MIMO integral feedback controller allowed us to stabilise the measured signals at set-points of zero, reducing sensitivity of the lock to changes in input power, insertion loss, or modulation depth.  We demonstrated carrier and negative sideband suppression of \SI{-26}{\deci\bel} relative to the desired positive sideband with the auxiliary tone at $\omegalf$ being \SI{-21}{\deci\bel} below the sideband.  CS-SSB operation was demonstrated with the lock engaged for $>\!\SI{e4}{\second}$ and under severe temperature changes.  We have already deployed our system for two atom interferometry experiments and have observed that the lock is maintained for weeks at a time even though the I/Q modulator is located next to the heating and cooling system for the laboratory.  Calculations predict that the systematic error resulting from the residual sidebands will be on the order of \SI{1}{\milli\radian}, which corresponds to a part-per-billion accuracy in the measurement of terrestrial gravity using a modest interferometer time of \SI{75}{\milli\second}.

Ultimately, suppression of the unwanted sidebands is limited by the I/Q modulator itself.  Imperfectly matched transmission losses between the phase modulators that comprise the two MZIs reduces carrier suppression, as do non-linearities in the phase modulator.  Suppression of the $-\omegahf$ sideband can be improved by independently controlling the phase and amplitude of the I and Q inputs to compensate for unwanted transmission losses and phase shifts \cite{kodigala_high-performance_2024}.  Imbalanced residual amplitude modulation (RAM) between phase modulators then sets a limit on the achievable suppression for the $-\omegahf$ sideband.  

The fast analog and digital control system presents opportunities for future improvements for reducing statistical and systematic errors.  Suppression of unwanted sidebands is likely limited by detection noise, but the signal-to-noise ratio of the $\omegalf$ measurement can be improved by using higher power in the $\omegalf$ driving signal.  Although increased power in the $+\omegalf$ sideband will increase the systematic error in atom interferometers, it is simple to implement a digital sample-and-hold on the control loop that pauses feedback and simultaneously disables the $\omegalf$ driving signals.  Given that the analog outputs have a bandwidth of \SI{50}{\mega\hertz}, and the main system clock is \SI{125}{\mega\hertz}, this sample-and-hold system can be switched on only during the Raman pulses with durations on the order of \SI{10}{\micro\second}.  Furthermore, after frequency-doubling, the $+\omegalf$ sideband (now at $+2\omegalf$ from the carrier) can be used to phase lock the I/Q modulator path and the carrier-only path using the same FPGA system, which will suppress low-frequency phase noise that occurs due to path length variations and improve interferometer sensitivity.  

\begin{backmatter}
\bmsection{Acknowledgments}
This work is supported by the Australia-India Strategic Research Fund (AISRF) grant no.~AIRXIV000025. RJT and SL were supported by Australian Research Council grants DP190101709 and LP19010062. SAH acknowledges support through an Australian Research Council Future Fellowship Grant No.~FT210100809.  The authors would like to thank Callum Sambridge and Jiri Janousek for their help with 1560 nm laser components. AU acknowledges fruitful discussions with Yosri Ben-A\"{i}cha, Rhys Eagle, and Ryan Husband.  RT acknowledges useful discussions with Ashok Kodigala.
\bmsection{Disclosures}

The authors declare no conflicts of interest.

\bmsection{Data Availability Statement}

Data underlying the results presented in this paper are not publicly available at this time but may be obtained from the authors upon reasonable request.

Hardware definition language (HDL) code for the FPGA architecture, as well as supporting software, is publicly available on GitHub \cite{github}.

\end{backmatter}


\begin{thebibliography}{10}
\newcommand{\enquote}[1]{``#1''}

\bibitem{kasevich_measurement_1992}
M.~Kasevich and S.~Chu, {\protect\JournalTitle{Applied Physics B}} \textbf{54},
  321 (1992).

\bibitem{peters_measurement_1999}
A.~Peters, K.~Y. Chung, and S.~Chu, {\protect\JournalTitle{Nature}}
  \textbf{400}, 849 (1999). Number: 6747 Publisher: Nature Publishing Group.

\bibitem{freier_mobile_2016}
C.~Freier, M.~Hauth, V.~Schkolnik, \emph{et~al.},
  {\protect\JournalTitle{Journal of Physics: Conference Series}} \textbf{723},
  012050 (2016).

\bibitem{bidel_absolute_2018}
Y.~Bidel, N.~Zahzam, C.~Blanchard, \emph{et~al.}, {\protect\JournalTitle{Nature
  Communications}} \textbf{9}, 627 (2018).

\bibitem{bidel_airborne_2023}
Y.~Bidel, N.~Zahzam, A.~Bresson, \emph{et~al.}, {\protect\JournalTitle{Journal
  of Geophysical Research: Solid Earth}} \textbf{128}, e2022JB025921 (2023).

\bibitem{phillips_storage_2001}
D.~F. Phillips, A.~Fleischhauer, A.~Mair, \emph{et~al.},
  {\protect\JournalTitle{Physical Review Letters}} \textbf{86}, 783 (2001).

\bibitem{appel_quantum_2008}
J.~Appel, E.~Figueroa, D.~Korystov, \emph{et~al.},
  {\protect\JournalTitle{Physical Review Letters}} \textbf{100}, 093602 (2008).

\bibitem{lvovsky_optical_2009}
A.~I. Lvovsky, B.~C. Sanders, and W.~Tittel, {\protect\JournalTitle{Nature
  Photonics}} \textbf{3}, 706 (2009).

\bibitem{hosseini_unconditional_2011}
M.~Hosseini, G.~Campbell, B.~M. Sparkes, \emph{et~al.},
  {\protect\JournalTitle{Nature Physics}} \textbf{7}, 794 (2011).

\bibitem{bluvstein_quantum_2022}
D.~Bluvstein, H.~Levine, G.~Semeghini, \emph{et~al.},
  {\protect\JournalTitle{Nature}} \textbf{604}, 451 (2022).

\bibitem{levine_dispersive_2022}
H.~Levine, D.~Bluvstein, A.~Keesling, \emph{et~al.},
  {\protect\JournalTitle{Physical Review A}} \textbf{105}, 032618 (2022).

\bibitem{bluvstein_logical_2024}
D.~Bluvstein, S.~J. Evered, A.~A. Geim, \emph{et~al.},
  {\protect\JournalTitle{Nature}} \textbf{626}, 58 (2024).

\bibitem{templier_carrier-suppressed_2021}
S.~Templier, J.~Hauden, P.~Cheiney, \emph{et~al.},
  {\protect\JournalTitle{Physical Review Applied}} \textbf{16}, 044018 (2021).

\bibitem{carraz_phase_2012}
O.~Carraz, R.~Charrière, M.~Cadoret, \emph{et~al.},
  {\protect\JournalTitle{Physical Review A}} \textbf{86}, 033605 (2012).

\bibitem{lee_compact_2022}
J.~Lee, R.~Ding, J.~Christensen, \emph{et~al.}, {\protect\JournalTitle{Nature
  Communications}} \textbf{13}, 5131 (2022). Number: 1 Publisher: Nature
  Publishing Group.

\bibitem{sane_11_2012}
S.~S. Sané, S.~Bennetts, J.~E. Debs, \emph{et~al.},
  {\protect\JournalTitle{Optics Express}} \textbf{20}, 8915 (2012).

\bibitem{kasevich_atomic_1991}
M.~Kasevich and S.~Chu, {\protect\JournalTitle{Physical Review Letters}}
  \textbf{67}, 181 (1991).

\bibitem{hosseini_high_2011}
M.~Hosseini, B.~M. Sparkes, G.~Campbell, \emph{et~al.},
  {\protect\JournalTitle{Nature Communications}} \textbf{2}, 174 (2011).

\bibitem{wang_field-deployable_2022}
Y.~Wang, A.~N. Craddock, R.~Sekelsky, \emph{et~al.},
  {\protect\JournalTitle{Physical Review Applied}} \textbf{18}, 044058 (2022).

\bibitem{menoret_gravity_2018}
V.~Ménoret, P.~Vermeulen, N.~Le~Moigne, \emph{et~al.},
  {\protect\JournalTitle{Scientific Reports}} \textbf{8}, 12300 (2018). Number:
  1 Publisher: Nature Publishing Group.

\bibitem{wu_gravity_2019}
X.~Wu, Z.~Pagel, B.~S. Malek, \emph{et~al.}, {\protect\JournalTitle{Science
  Advances}} \textbf{5}, eaax0800 (2019). Publisher: American Association for
  the Advancement of Science Section: Research Article.

\bibitem{kodigala_high-performance_2024}
A.~Kodigala, M.~Gehl, G.~W. Hoth, \emph{et~al.}, {\protect\JournalTitle{Science
  Advances}} \textbf{10}, eade4454 (2024).

\bibitem{theron_narrow_2015}
F.~Theron, O.~Carraz, G.~Renon, \emph{et~al.}, {\protect\JournalTitle{Applied
  Physics B}} \textbf{118}, 1 (2015).

\bibitem{li_novel_2020}
R.~Li, X.~Sun, and D.~Yang, {\protect\JournalTitle{IEEE Photonics Technology
  Letters}} \textbf{32}, 815 (2020).

\bibitem{izutsu_integrated_1981}
M.~Izutsu, S.~Shikama, and T.~Sueta, {\protect\JournalTitle{IEEE Journal of
  Quantum Electronics}} \textbf{17}, 2225 (1981).

\bibitem{hall_photorefractive_1985}
T.~J. Hall, R.~Jaura, L.~M. Connors, and P.~D. Foote,
  {\protect\JournalTitle{Progress in Quantum Electronics}} \textbf{10}, 77
  (1985).

\bibitem{kong_recent_2012}
Y.~Kong, S.~Liu, and J.~Xu, {\protect\JournalTitle{Materials}} \textbf{5}, 1954
  (2012). Number: 10 Publisher: Molecular Diversity Preservation International.

\bibitem{bui_instrumentation_2011}
D.~T. Bui, C.~T. Nguyen, I.~Ledoux-Rak, \emph{et~al.},
  {\protect\JournalTitle{Measurement Science and Technology}} \textbf{22},
  125105 (2011).

\bibitem{wald_analog_2023}
S.~Wald, F.~Diorico, and O.~Hosten, {\protect\JournalTitle{Applied Optics}}
  \textbf{62}, 1 (2023).

\bibitem{github}
R.~Thomas and A.~Ullah, \enquote{{VHDL and MATLAB code for I/Q bias
  stabilisation},} https://github.com/atomlaser-lab/iq-bias-control (2024).

\bibitem{freier_atom_2017}
C.~Freier, \enquote{Atom interferometry at geodetic observatories,} Ph.D.
  thesis (2017). Accepted: 2017-06-18T16:04:41Z Publisher:
  Humboldt-Universität zu Berlin, Mathematisch-Naturwissenschaftliche
  Fakultät.

\end{thebibliography}
\end{document}